\documentclass{WileyMSP-template}

\usepackage{mhchem}
\usepackage{pdfpages}

\newcommand{\dC}{$^\circ$C}
\newcommand{\diele}{$\varepsilon_{\rm poly}^\infty$}
\newcommand{\dielt}{$\varepsilon_{\rm poly}^0$}
\newcommand{\etal}{\textit{et al}.}
\newcommand{\Ehull}{E_{\rm hull}}

\begin{document}

\pagestyle{fancy}
\rhead{\includegraphics[width=2.5cm]{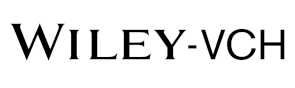}}

\title{Screening 0D materials for 2D nanoelectronics applications}

\maketitle


\author{Mohammad Bagheri}
\author{Hannu-Pekka Komsa*}

\begin{affiliations}
M. Bagheri, Prof. H.-P. Komsa \\
Microelectronics Research Unit, University of Oulu,
P.O. Box 8000, 90014 Oulu, Finland
\\
Email Address: hannu-pekka.komsa@oulu.fi
\end{affiliations}


\begin{abstract}
  As nanoelectronic devices based on two-dimensional (2D) materials are moving towards maturity, optimization of the properties of the active 2D material must be accompanied by equal attention to optimizing the properties of and the interfaces to the other materials around it, such as electrodes, gate dielectrics, and the substrate. While these are usually either 2D or 3D materials, recently K. Liu \etal [Nat. Electron. 4, 906 (2021)] reported on the use of zero-dimensional (0D) material, consisting of vdW-bonded Sb$_2$O$_3$ clusters, as a highly promising insulating substrate and gate dielectric. Here, we report on computational screening study to find promising 0D materials for use in nanoelectronics applications, in conjunction with 2D materials in particular. By combining a database and literature searches, we found 16 materials belonging to 6 structural prototypes with high melting points and high band gaps, and a range of static dielectric constants. We carried out additional first-principles calculations to evaluate selected technologically relevant material properties, and confirmed that all these materials are van der Waals-bonded, thus allowing for facile separation of 0D clusters from the 3D host and also weakly perturbing the electronic properties of the 2D material after deposition.
\end{abstract}


\section{Introduction}

Two-dimensional (2D) materials are widely investigated
for use in nanoelectronics devices,
especially as an ultrathin channel in next-generation
field-effect transistors and novel logic architectures
\cite{Radisavljevic11_NNano,Liu19_NNano,Marega20_Nat},
memory cells \cite{Bertolazzi13_ACSNano,Liu18_NNano},
memristors \cite{Sangwan15_NNano,Ge18_NL}, rectenna \cite{Zhang19_Nat},
and others \cite{Liu20_NNano}.
In all of these applications, one needs to optimize not only
the properties of the ``active'' 2D material, but also all the other
materials around it,
such as the dielectrics, electrodes, and substrate, and the
material interfaces between them. 
In a wide majority of cases, either other 2D materials or
common 3D materials are used for these purposes.
Recently, formation of 2D molecular crystal was demonstrated from
\ce{Sb2O3} 0D material, i.e., a material consisting of weakly bound
small molecular units with no dangling bonds \cite{Han19_NComm}.
Using these as a substrate or gate dielectric in 2D material
field-effect transistors resulted in high mobility, small hysteresis,
low density of trap states, and high subthreshold slope \cite{Liu21_NEle}.
The authors called for theoretical studies to find more suitable
0D material candidates \cite{Liu21_NEle,Liu22_JPCL}.

While 2D and 1D materials have been searched in many previous
publications by screening experimental and/or computational material databases
\cite{Lebegue13_PRX,Cheon17_NL,Mounet18_NNano,Friedrich22_NL},
to the best of our knowledge, not 0D materials.
While 0D materials are easily identified using the same dimensionality identification
methods, many of the found materials are crystalline phases
of substances that are gaseous or liquid even in atmospheric conditions,
and thus not useable in the above-mentioned applications.
Finding the materials that are crystalline at technologically relevant
temperatures requires estimating the melting points, which
is computationally highly demanding.

In this Letter, we take first steps in identifying a set of
promising 0D materials for use in electronics applications,
as a substrate or as a gate dielectric.
We achieve this by combining screening of the Materials Project
database and supplementing it by manual pruning of the list,
most importantly by extracting experimental melting points
from the literature.
We identified 16 materials belonging to 6 structural prototypes
with high melting points 
and high band gaps, and both high and low dielectric constants.
We performed van der Waals density-functional theory calculations
to evaluate selected material properties
and also studied their interaction with the prototypical 2D material MoS$_2$.

\section{Results}

\subsection{Database screening}

\begin{figure*}[!ht]
\begin{center}
  \includegraphics[width=16cm]{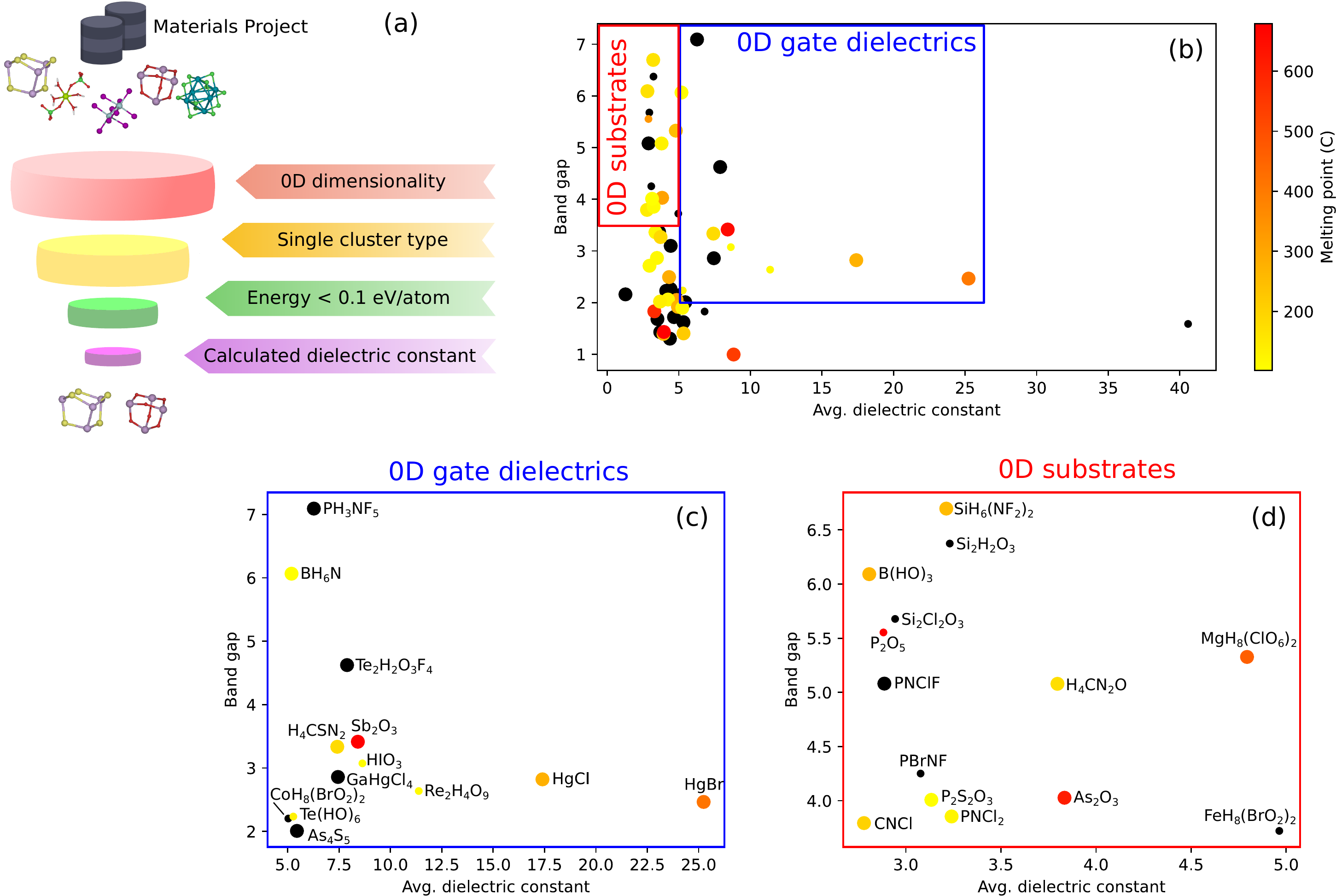}
\end{center}
\caption{\label{fig:gapdiel}
  (a) Schematic representation of the materials database screening procedure.
  (b-d) Band gap vs.\ average static dielectric constant from
  (b) all 56 materials with melting poit above 100 \dC,
  (c) materials with band gap larger than 2 and dielectric constant larger than 5, and
  (d) materials with band gap larger than 3.5 and dielectric constant less than 5.
  Marker color indicates the melting point (see colorbar, or black if
  not known). Large/small marker size indicate that the material
  is stable/unstable, defined as the energy above convex hull being
  less/more than 10 meV/atom.
}
\end{figure*}

We start by performing a database search to Materials Project (MP) \cite{MP},
as illustrated in Fig.~\ref{fig:gapdiel}(a),
with the following search criteria
\begin{itemize}
\item[Cr1] The structure is classified to have 0D dimensionality.
\item[Cr2] The structure consist of only one type of 0D cluster
  and the size of the cluster is larger than 3 atoms.
\item[Cr3] Energy above convex hull $\Ehull$ is less than 0.1 eV/atom.
\item[Cr4] The MP entry contains calculated dielectric constant.
\end{itemize}

Several approaches have been proposed for 
the classification of structure dimensionality
\cite{Gorai16_JMCA,Cheon17_NL,Larsen19_PRM}.
We are using the algorithm proposed by Larsen \etal{}~\cite{Larsen19_PRM},
which also gives a useful measure of the degree of ``0D-ness''.
Upon applying Cr1,
out of 126335 materials in MP (accessed on 25th Jan 2022),
8335 were identified as 0D materials.

Criterion Cr2 is chosen for simplicity and since deposition of materials consisting
of many different clusters can be difficult to control,
especially since the deposition is ideally carried out at a relatively
low temperature.
The clusters are compared only by the formula
(number and types of atoms in the cluster), not their atomic structures.
Application of this criterion (on top of Cr1) yields 2395 materials.
%
In addition, we have required that the number of atoms in the cluster
is larger than 3.
Clusters with one atom can be trivially ignored (mostly noble gases),
and most systems with 2- or 3-atom clusters correspond to solid phases of,
e.g., hydrogen halides (HF, HCl, etc.) or
simple molecules such as \ce{N2}, \ce{H2O}, and \ce{CO2}.
It is unlikely that very promising new 0D materials can be found among these.
Finally, we have required that the ``0D-ness'' dimensionality score
used in the classification scheme is higher than 0.8 \cite{Larsen19_PRM}.
The threshold was found by manually inspecting the structures
with lowest score and comparing the dimensionality to that
reported in MP. The materials with low dimensionality score
are borderline between different dimensionalities and usually
indicates that the bonding between the (nearly) 0D clusters is
stronger than in purely vdW-bonded systems.
Application of these criteria yields 1965 materials.

Applying the energy criterion Cr3, reduces the number of
material candidates by about a factor of two to 942.

Finally, in the last step of our database screening,
criterion Cr4 yields a marked reduction in the number of
entries, down to 162. These materials are listed in Table S1.
Omitting such a large number of entries (780) may appear unwarranted,
but we will revisit this group of materials later in this paper.

In the next stage, we manually pruned the list. 
First, if several phases of same material are found,
we retain only the lowest energy phase, down to 138 materials.
Inspection of the list also indicates that many of these materials
are liquid or gaseous under atmospheric conditions (e.g., methane, ammonia,
and various acids).
In order to find materials with stable crystalline structure at
moderate processing temperatures,
we collected experimentally measured melting points
for all materials where we could find one (also listed in Table S1).
After removing the materials that are known to be liquid or gas
under typical operating temperatures (melting point below 100 \dC),
we are left with 56 entries.

The band gaps with respect to rotationally averaged
static dielectric constants,
or ``polycrystalline dielectric constant'', \dielt{}
for these 56 materials are plotted in
Fig.~\ref{fig:gapdiel}(b).
In electronics applications with relatively slowly varying
fields, the static dielectric constants are more relevant
than the high-frequency one $\varepsilon_\infty$.
In fact, the ionic contribution to the dielectric constant is critical
for achieving large screening, as evidenced from the plot of the band
gap vs.\ $\varepsilon_\infty$ shown in Figure S1.

\begin{figure*}[!ht]
\begin{center}
  \includegraphics[width=12cm]{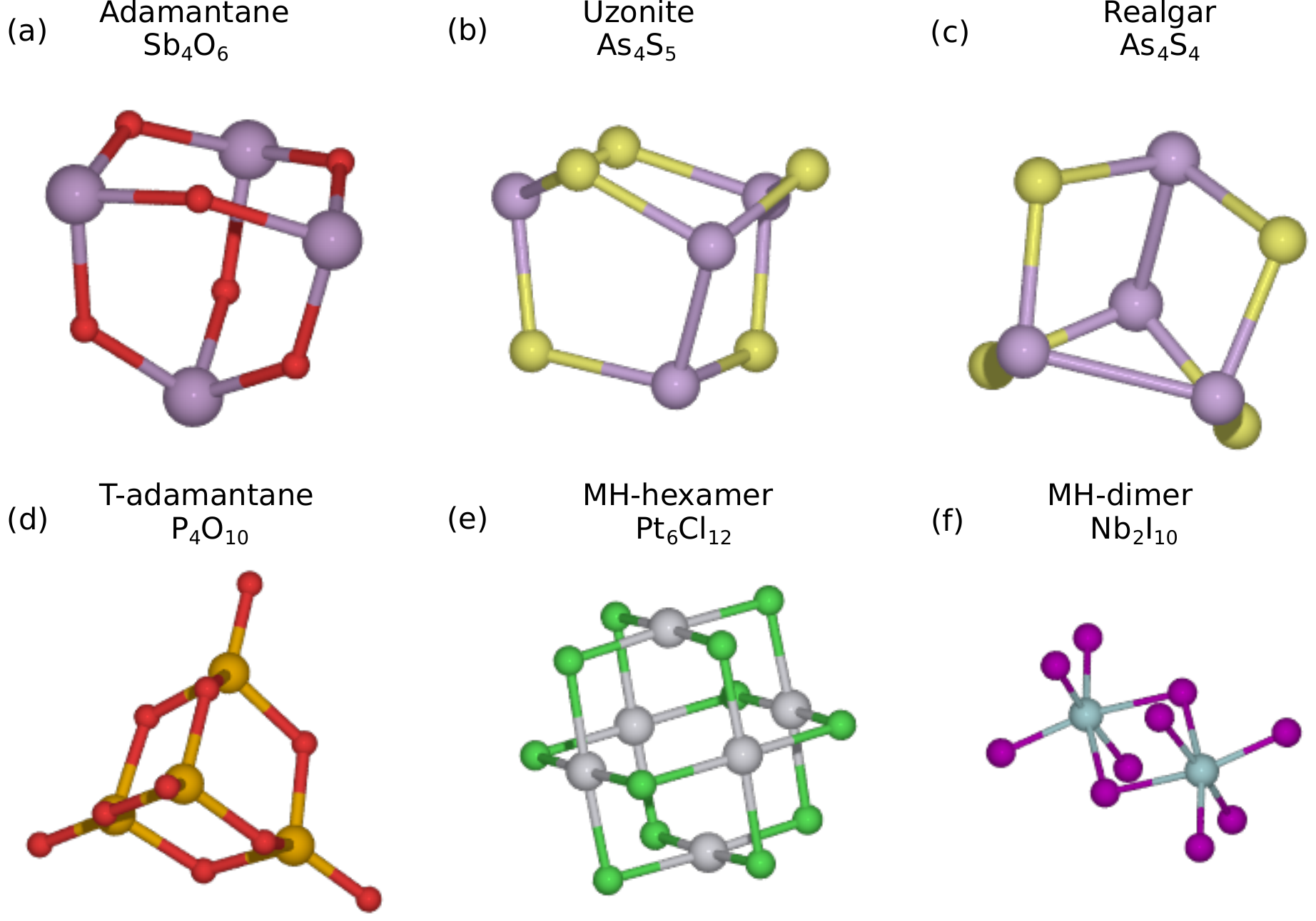}
\end{center}
\caption{\label{fig:geom}
  Atomic structures of the cluster prototypes.
  }
\end{figure*}

In order to find materials suitable for use as gate dielectrics,
we extract materials with band gap larger than 2 eV and
\dielt{} larger than 5.
The relatively low band-gap threshold was chosen since DFT with
semilocal functionals underestimates gaps, sometimes significantly so,
by upto a factor of two.
In this group, shown in Fig.~\ref{fig:gapdiel}(b),
there are only 13 materials, which we analyze in detail.
The atomic structures of the corresponding 0D clusters are shown in Fig. S2
(and in Fig.~\ref{fig:geom} for the promising candidates).

The only material with very high melting point (mp.) is \ce{Sb2O3}
(656 \dC), that was already previously identified \cite{Han19_NComm,Liu21_NEle}.
The atomic structure consists of \ce{Sb4O6} cages,
similar to \ce{C10H16} carbon cages, called adamantane, but without terminating H.
There are few materials with higher dielectric constant, such as
\ce{HgBr}, \ce{HgCl}, and \ce{Re2H4O9}, but those have also lower band gap and
lower melting point. 
Moreover, although \ce{HgBr} and \ce{HgCl} have fairly high mp.,
they are unfortunately highly toxic.
\ce{Re2H4O9} structure is similar to, but not exactly the same as,
that of perrhenic acid.
It is a metal oxide coordinated with water with mp. of 115 \dC, which
is likely too low.

Among others (from high to low band gap),
\ce{PH3NF5} is a Lewis acid-base adduct, with fairly little experimental data.
It has been synthesized, but expected to have low melting point
\cite{Storzer83_CB}.
\ce{BH6N} has a low mp. of 104 \dC.
No mp. was found for \ce{Te2H2O3F4}, but stable at room temperature and computationally stable.
\ce{H4CSN2}, thiourea, has mp. of 182 \dC, making it possibly interesting.
\ce{HIO3}, iodic acid, has band gap and dielectric constant close
to \ce{Sb2O3}, but the mp. of 110 \dC{} is again low.
\ce{GaHgCl4} has been synthesized and is computationally stable. No mp. was found and it is sensitive to humidity \cite{Rosdahl04_ZAAC}.
\ce{CoH8(BrO2)2} (or \ce{CoBr2(H2O)4}) consist of hydrated \ce{CoBr2} molecules.
The degree of hydration is sensitive to temperature between 0-100 \dC \cite{Benrath38_ZAAC}, which may prove problematic in applications.
We could not find mp., but high $\Ehull$ of 0.053 eV/atom
suggests poor stability.
\ce{Te(HO)6}, telluric acid, has mp. of 136 \dC, which is likely too low.
Finally, \ce{As4S5} is a rare mineral called uzonite \cite{Whitfield73_JCSDT}.
It is related to another mineral \ce{As4S4}, realgar,
also with fairly similar cage-like structure, and mp. 320 of \dC.
Thus, although the band gap and dielectric constant are fairly low,
the cage-like structure and stability of arsenic sulfides makes
them promising.

Eventually, the number of viable candidates found by this screening
is small. In addition to \ce{Sb2O3}, there are only 
\ce{As4S5} (and maybe very similar \ce{As4S4}) (uzonite/realgar).
The commonness of thiourea could make it interesting in some cases.

For use as a substrate, the band gap is often more important
than the dielectric constant.
Low dielectric constant materials (low-$\kappa$ dielectrics)
are also used in separating the interconnects in circuits,
o reduce crosstalk and parasitic capacitance.
Moreover, in optoelectronic applications low dielectric constant might be desired in order to minimize sreening by the substrate and to avoid subsequent
renormalization of fundamental band gap and/or reduction in exciton binding energy
\cite{Komsa12_PRB2,Lin14_NL,Chernikov14_PRL,Ugeda14_NMat}.
We next focus on the large band gap ($>3.5$ eV) but small dielectric
constant materials \dielt$<5$, which yields 14 materials
shown in Figure \ref{fig:gapdiel}(d)
and the structures in Fig.~S3.
Two interesting candidates can be immediately
pinpointed: \ce{As2O3} and \ce{P2O5}.
\ce{As2O3} has the same structure as \ce{Sb2O3}, i.e., belonging to adamantanes.
It has mp. of 312 \dC{} and is also found as a mineral arsenolite
(and in small quantities in arsenic sulfide minerals uzonite and realgar
mentioned above).
\ce{P2O5} is also reminiscent of adamantanes but with additional
terminating oxo groups. It has mp. 340 \dC, and while this polymorph
consisting of \ce{P4O10} molecules is in principle metastable,
it is nevertheless commonly encountered.
We note that \ce{P2S2O3} consists of \ce{P4O6S4} clusters
similar to \ce{P2O5} but the tip O replaced by S.
However, it has much lower mp. of 102 \dC.

Highest band gaps are found from silicon clusters.
\ce{SiH6(NF2)2} is an adduct of \ce{SiF4} and 2\ce{NH3} and computatioally
stable, although mp. of only 166 \dC.
\ce{Si2H2O3} and \ce{Si2Cl2O3} cluster both have pyramidal structure
consisting of four formula units, similar to \ce{P4O10}.
We did not find mp., but $\Ehull=0.023$ and $\Ehull=0.026$ eV/atom, respectively,
suggests fairly poor stability.
Highest dielectric constants are found from metal halogen compounds.
\ce{FeH8(BrO2)2} has a structure similar to \ce{CoH8(BrO2)2} \cite{Waizumi92_ICA},
and is similarly sensitive to humidity and exhibits fairly
poor stability with $\Ehull=0.026$ meV/atom.
\ce{MgH8(ClO6)2} (or \ce{Mg(ClO4)2(H2O)4}) is a hydrated version of magnesium perchlorate, which is a strong drying agent, which may again prove
problematic. On the other hand, it has fairly high mp. of 251 \dC and computationally stable.

Otherwise,
\ce{B(HO)3}, boric acid, has mp. 171 \dC, and is also found in mineral sassolite, but dissolves in water.
\ce{H4CN2O}, urea, has mp. of only 133 \dC{} and thiourea has overall better properties.
\ce{PNCl2} is phosphazene with 6-membered ring with alternating P and N,
but mp. of 112 \dC{} is too low.
\ce{PNClF} and \ce{PBrNF} are similar and, though they are stable at RT, the mp. is likely low as in other phosphorus halides \cite{Clare74_JCSDT}.
\ce{CNCl}, cyanuric chloride, consists of 6-membered ring with
alternating C and N, and has mp. of 144 \dC.

Thus, from this set, we found only two clearly promising materials
to be used as a substrate: \ce{As2O3}, \ce{P2O5}.
Possibly interesting candidates are magnesium perchlorate and boric acid.

Finally, it is worth highlighting few specific data points in Fig.~\ref{fig:gapdiel}(b).
The material with dielectric constant over 40 is \ce{H2Pt(OH)6}, platinic acid.
In crystalline form, it is experimentally stable to about 150 \dC, after
which it starts to form \ce{PtO2} \cite{Venediktov12_RJAC}.
The calculated $\Ehull=0.069$ eV/atom also suggests poor stability.
There are also three high melting point materials, shown by red markers,
in Fig.~\ref{fig:gapdiel}(b): \ce{PtCl2}, \ce{PdCl2}, and \ce{NbI5}.
The Pt and Pd chlorides in $\beta$-phase consist of octahedral cage structure
with six formula units (hexamer). \ce{PtCl2} and \ce{PdCl2} are reported to have high
mp. of 581 and 679 \dC, respectively, but those values correspond to
the polymeric $\alpha$-phase.
There is a phase transition from $\beta$- to $\alpha$-phase at around 500 \dC \cite{Krebs88_ZAAC}, which is still high.
The calculated band gaps are less than 2 eV and dielectric constant 3--4.
\ce{NbI5} is a molecular dimer with mp. 543 \dC,
gap 1.0 eV, and \dielt{}$=8.8$.

From this database search, we have ended up with 8 materials
in 6 cluster prototypes that showed high melting points and
reasonably high band gap and dielectric constant:
adamantane (\ce{Sb4O6}, \ce{As4O6}),
uzonite (\ce{As4S5}),
realgar (\ce{As4S4})
terminated adamantane (denoted T-adamantane, \ce{P4O10}),
metal halide hexamer (MH-hexamer, \ce{Pt6Cl12}, \ce{Pd6Cl12}),
and metal halide dimer (MH-dimer, \ce{Nb2I10}).
The prototype cluster structures are visualized in Fig.~\ref{fig:geom}.

We now revisit the materials for which calculated dielectric constant
was not available in MP.
We represented all cluster structures as connected graphs and then searched
for graphs that are isomorphic to the prototype systems
among those where dielectric constant was not available.
We stress, that this approach only considers 
the connectivity/bonding network between atoms, but
ignores the atom types and thus, for example, \ce{P4O10} and \ce{P4S4O6} are isomorphic
despite having only a fraction of the oxygens replaced by sulfur.
This search yielded in total 27 structures,
3 adamantanes, 1 uzonite-like, 2 realgar-like, 3 T-adamantanes,
0 MH-hexamer, 18 MH-dimer. These are listed in Table S2.
We again pruned the list by removing low mp. and metastable
systems, and those with zero or vanishingly small band gap.
It is worth noting that four of the metal halide dimers with zero band gap
also exhibited nonzero magnetic moment.
We are left with 12 additional materials:
uzonite-like \ce{As4Se4} with mp. of 265 \dC;
T-adamantane \ce{P4S10}, \ce{Ge4S6Br4}, \ce{Ge4S6I4} all have mp close to 300 \dC;
MH-dimers \ce{Nb2Br10} and \ce{Ta2Br10} have mp. 250--300 \dC,
\ce{Nb2Cl10} has mp. close to 200 \dC, and \ce{Ta2I10} has mp. of 382 \dC.

\subsection{Cluster properties}

From the screening study presented thus far,
we are left with 16 materials in the 6 prototypes.
Since Materials Project data is calculated using PBE functional
that poorly describes van der Waals interactions, we recalculated
these systems, both crystalline and isolated clusters,
using vdW-DF functional. 
Selected properties are collected in Table \ref{tab:cands},
such as binding energies, band gaps and dielectric constants,
and additional properties are listed in Table S3.
The experimental melting point is also included for convenience
in Table \ref{tab:cands}.

The binding energies per cluster are usually few eVs, which
may appear quite large, but the sizes of these clusters and
the areas between them are also significant.
We estimate binding energy per area by mapping the volume per molecule in a crystal to a sphere and taking its area.
These values fall within 6--40 meV/{\AA}, which is comparable to
the 2D material binding energies, where majority of compounds
fall within 13--21 meV/{\AA} \cite{Bjorkman12_PRL,Bjorkman14_JCP}.
Thus, it appears safe to label these systems as vdW-bonded
0D materials.


\begin{table*}[!ht]
  \caption{\label{tab:cands}
    Selected properties of promising 0D materials calculated
    using vdW-DF functional: cluster formula,
    binding energy per area (meV/{\AA}), band gap from crystal (eV),
    HOMO-LUMO gap from cluster (eV), and static and high-frequency
    dielectric constants.
    The experimental melting points (mp., in \dC) are also included
    for convenience.
  } 
\begin{tabular}{lcccccccc}
formula & binding energy & mp. & Crystal gap & Cluster gap & \diele & \dielt \\
\ce{Sb4O6} & -30.37 & 656.0 & 2.92 & 4.09 & 3.95 & 10.92 \\
\ce{As4S5} & -18.66 &  & 1.42 & 2.64 & 6.31 & 8.10 \\
\ce{As4S4} & -17.29 & 320.0 & 1.38 & 2.80 & 6.80 & 7.91 \\
\ce{As4O6} & -18.30 & 312.2 & 4.04 & 4.49 & 2.59 & 4.92 \\
\ce{P4O10} & -13.40 & 340.0 & 5.47 & 6.01 & 1.37 & 2.41 \\
\ce{Pt6Cl12} & -6.15 & 581.0 & 1.48 & 2.01 & 3.99 & 4.19 \\
\ce{Pd6Cl12} & -6.50 & 679.0 & 1.09 & 1.64 & 6.25 & 6.80 \\
\ce{Nb2I10} & -8.04 & 543.0 & 0.61 & 1.08 & 9.98 & 15.01 \\
\ce{As4Se4} & -39.87 & 265.0 & 0.85 & 2.17 & 10.69 & 12.61 \\
\ce{P4S10} & -12.33 & 288.0 & 2.17 & 2.97 & 3.98 & 4.60 \\
\ce{Ge4S6Br4} & -11.46 & 305.0 & 2.14 & 2.79 & 3.40 & 4.22 \\
\ce{Ge4S6I4} & -11.02 & 310.0 & 2.06 & 2.59 & 3.85 & 4.75 \\
\ce{Nb2Br10} & -26.47 & 254.0 & 1.37 & 1.70 & 4.72 & 7.13 \\
\ce{Ta2Br10} & -18.94 & 265.0 & 1.86 & 2.19 & 3.85 & 5.79 \\
\ce{Nb2Cl10} & -17.27 & 204.7 & 2.07 & 2.33 & 3.12 & 4.99 \\
\ce{Ta2I10} & -30.31 & 382.0 & 0.94 & 1.41 & 7.25 & 10.54 \\

\end{tabular}
\end{table*}

\begin{figure*}[!ht]
\begin{center}
  \includegraphics[width=16cm]{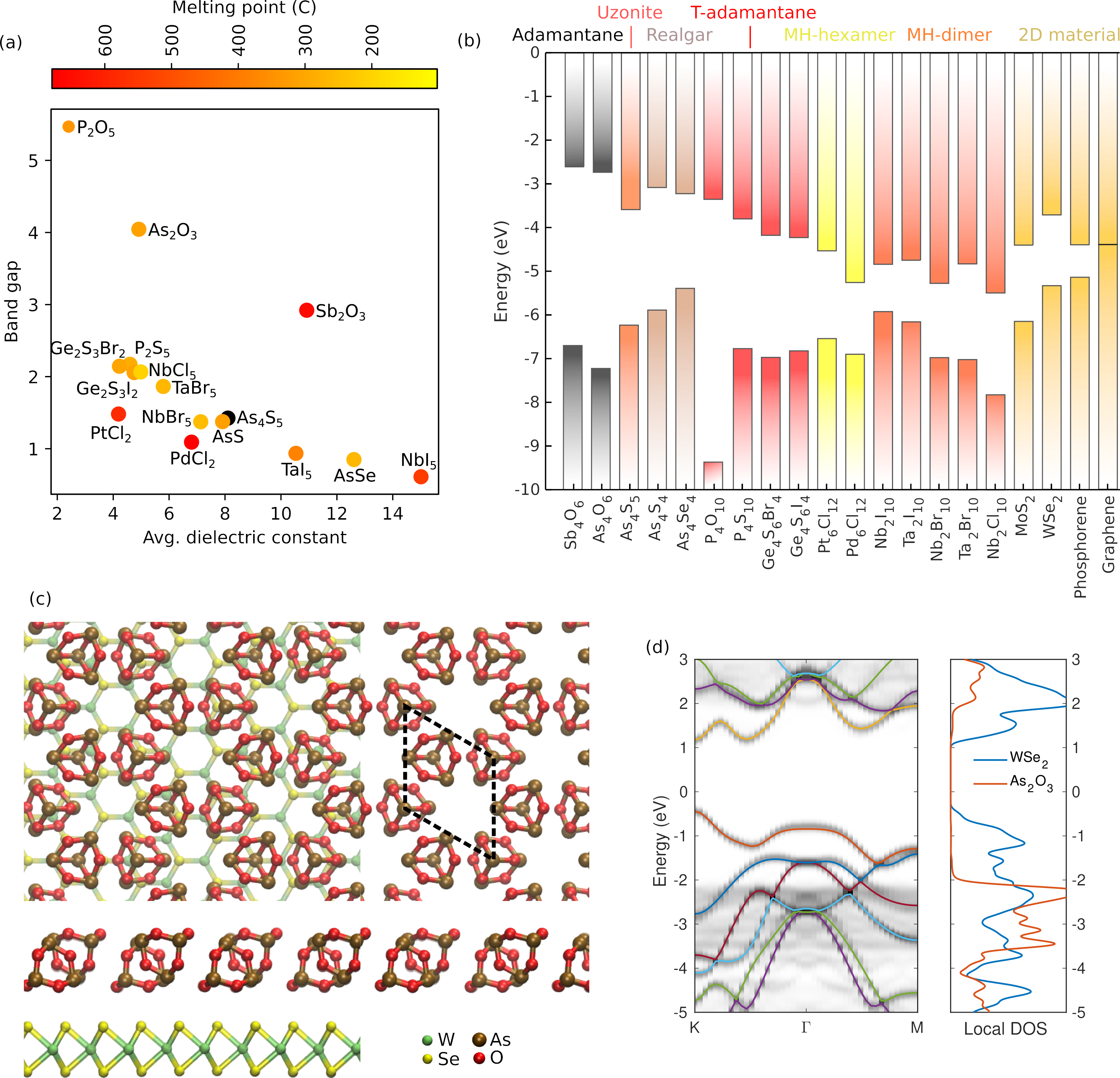}
\end{center}
\caption{\label{fig:props}
  Properties from vdW-DF calculations.
  (a) Band gap vs.\ static dielectric constant.
  Marker color indicates the melting point or black if unknown.
  (b) HOMO and LUMO levels of 0D clusters aligned via vacuum level.
  The materials are grouped and colored according to the cluster
  prototype.
  (c) Top and side vies of atomic structure of \ce{As2O3} on \ce{WSe2}.
  Dashed rhombus denotes the \ce{As2O3} surface unit cell.
  (d) Effective band structure of the heterostructure model
  (underlying gray features)
  compared to the band structure of pristine \ce{WSe2} (overlaid lines).
  The local DOS of \ce{As2O3} and \ce{WSe2} layers are also shown.  
  }
\end{figure*}

In Table \ref{tab:cands}, we also listed the gaps calculated using
vdW-DF for crystals and for isolated clusters.
Comparing to the MP gaps listed in Tables S1 and S2, shows that,
as van der Waals interaction pulls the clusters closer together, the band gap
decreases: 0.4--0.5 eV decrement is seen for many materials, except for
\ce{As2O3} and \ce{P2O5} which remain largely unaffected.
The lowering of the band gap is in turn reflected in slight increase of
the dielectric constants,
see also the plot of band gap vs.\ dielectric constant shown in Fig.~\ref{fig:props}(a).
Conversely, the HOMO-LUMO gaps of the isolated clusters are higher than
the crystal gaps by about 0.5--1 eV.


For nanoelectronics applications, it is important to know the alignment
of the valence band maxima and conduction band minima of the 0D materials
with respect to the band extrema of the 2D materials.
Here, we evaluated this from the HOMO and LUMO levels of isolated clusters
and the results are shown in Fig.~\ref{fig:props}(b).
Adamantanes, uzonite, realgars, and some of T-adamantanes have all
suitable band edge positions compared to common 2D materials.
Metal halide hexamer and dimer band edges are deep and as a result
the LUMO level of 0D cluster are below the conduction band minima of most
2D materials.

0D materials have several natural cleaving planes, that lead to 2D layers
of clusters with different orientations and different lattices,
depending on the symmetry of the clusters and of the 3D lattice.
It is not {\it a priori} clear what will be the preferred orientation and
ordering of the clusters when sputtered on top of a particular 2D material
and how to control the growth mode \cite{Liu22_JPCL}.
Investigating this by first-principles calculations is highly demanding.

To shed light on the interaction of the clusters with
2D materials, we first studied the interaction of isolated 0D clusters placed
on a monolayer \ce{MoS2}.
The density of states for all 16 cases are collected in Fig.~S4
and shows that the \ce{MoS2} states are largely unaffected by the 0D cluster,
evidencing that the interaction is weak.
Moreover, in Fig.~S4 we also indicate the binding energies.
For the interaction area we have again transformed the volume per cluster
to a sphere (as mentioned above) and taken its great circle.
The values fall within 10--20 meV/{\AA}, again similar to
those reported above and as reported for many 2D materials, providing
evidence to the vdW-type binding with the 2D material.

Next, as an example, we constructed one heterostructure with
the surface of \ce{WSe2} fully covered by \ce{As2O3}.
Due to the similarity of \ce{As2O3} and \ce{Sb2O3}, 
the structure of 0D clusters on the surface is also likely similar,
and thus we can adopt for \ce{As2O3} the experimentally determined
structure of \ce{Sb2O3}.
For the 2D material we chose \ce{WSe2} due to suitable lattice constant,
which allowed us to construct heterostructure model with minimal strain.
\ce{As2O3} has higher band gap than \ce{Sb2O3} but smaller dielectric
constant and thus useful as a low-$\kappa$ dielectric or,
e.g., as a substrate for studying
single-photon emitters in WSe$_2$ \cite{Chakraborty15_NNano,Toth19_ARPC}
without interference from the substrate defects.
The optimized atomic structure is shown in Fig.~\ref{fig:props}(c).
The atomic structures of both the \ce{WSe2} and \ce{As2O3} layers are
very weakly perturbed by the neighboring layer,
indicating again van der Waals type interaction.
Similarity of the effective band structure of the heterostructure
and the band structure of pristine \ce{WSe2} in Fig.~\ref{fig:props}(d)
demonstrates that the band structures of \ce{As2O3} and \ce{WSe2}
are effectively uncoupled.
The \ce{As2O3} bands with very small dispersion
are seen at about 2 eV below valence band maximum
and about 1 eV above conduction band minimum  of \ce{WSe2}.
Thus, charge transfer to/from the \ce{As2O3} should not occur
and we expect the electronic and optical properties of \ce{WSe2} layers
to remain largely intact.

\section{Conclusion}

In conclusion, we have extracted a list of promising 0D materials
to be used with 2D materials in nanoelectronics applications,
exhibiting fairly high band gaps, dielectric constants, and melting points.
This was achieved by screening Materials Project database, combining it
with experimental melting points, and using graph-theoretical
methods to find other similar structures.
We used first-principles calculations to predict additional material properties
and studied the interaction of the 0D clusters with a prototypical 2D materials MoS$_2$.
All these materials were confirmed to have binding energies typical
for vdW-bonded systems and to weakly perturb the electronic structure
of the 2D material.

Owing to the vdW gaps between clusters, their conductivity is expected to be
low and thus they should be fairly good insulators even though their band gaps in
the crystalline form are not very high.
Consequently, ``metallic'' 0D materials do not appear particularly promising
as electrode materials.
In addition to their use as a gate dielectric or substrate,
these 0D materials could be used as spacer layers in 2D material superlattices \cite{Kumar22_NNano}.
Intercalation between 2D layers can also stabilize crystalline order for clusters
that have too low melting point in 3D lattice.
For instance, superlattices of 2D materials and CTAB (and other related) molecules
was demonstrated in \cite{Wang18_Nat}.
Also, \ce{AlCl3} and \ce{CuCl2} clusters
and various dimerized and trimerized forms of them have successfully been
intercalated between graphene sheets \cite{Lin21_AM}.
Although materials consisting of more than one type of clusters were
ignored in this Letter for reasons of simplicity, a vast variety of
such 0D molecular compounds could exist.
As an example, 2D crystal made of \ce{SbI3} and \ce{S8} molecular units
was reported in Ref.~\cite{Feng20_AM}.


We hope that the our study motivates further experimental
and computational work on these materials.
Based on the cluster prototypes identified here, 
other 0D materials could be identified or purposefully designed,
either by substitution or by addition of functional groups.
These can extend the range of properties or provide even new functionalities,
such as charge doping or magnetism.

\section{Methods}

The material information were extracted from Materials Project
we used the REST API \cite{MPAPI}.
The structures were then fed into the dimensionality analysis
tools as implemented in ASE \cite{ase}.
The graph-theoretical part was done using NetWorkX package \cite{networkx}.

All density-functional theory calculations were carried out
using VASP software \cite{vasp1,vasp2},
together with projector augmented plane wave method. 
We adopted Hamada's rev-VDW-DF2 functional \cite{Hamada14_PRB,Dion04_PRL,RomanPerez09_PRL,Klimes11_PRB},
which was found to yield good structural properties and
interlayer binding energies for 2D materials \cite{Bjorkman14_JCP}.
Most other input parameters were taken from the Materials Project
\cite{MP,MPAPI}, such as the k-point meshes and the plane wave cutoff of 520 eV.
Isolated molecules on \ce{MoS2} were modeled
using 4$\times$4 supercell of \ce{MoS2} and a 3$\times$3 k-point mesh.
For the heterostructure of \ce{As2O3} and \ce{WSe2},
we take the [111] plane of \ce{As2O3}.
This has lattice constant of 7.715 {\AA}, in which case
a 3$\times$3 supercell ($23.144$ {\AA}) matches closely with
a 7$\times$7 supercell of \ce{WSe2} ($7 \cdot 3.297=23.079$ {\AA}). In the heterostructure model, we keep the lattice constant of \ce{WSe2} fixed and compressively strain \ce{As2O3} layer by 0.28\%.
We used 2$\times$2 k-point mesh during relaxation and
4$\times$4 k-point mesh in calculating density of states.
Effective band structure was calculated using BandUP code \cite{Bandup1}.

\medskip
\textbf{Supporting Information} \par 
Supporting Information is available from the Wiley Online Library or from the author.

\medskip
\textbf{Acknowledgements} \par 
We thank CSC--IT Center for Science Ltd. for generous grants of computer time.

\medskip
\textbf{Conflict of Interest} \par 
The authors declare no conflict of interest.

\medskip

%
\bibliographystyle{MSP}
\bibliography{0d,../../../../../bibbase,../../../../../publist}



\section*{Supplementary material}

# Screening 0D materials for 2D nanoelectronics applications

Mohammad Bagheri and Hannu-Pekka Komsa

[1]*Microelectronics Research Unit, University of Oulu, Oulu, Finland*

(Dated: March 15, 2022)

## I. TABLES

TABLE I: List of of all 162 materials found from the MP database search. Formula, MP-ID, $E_{\text{hull}}$ (in eV), $E_g$ (in eV), and $\varepsilon_0$ are extracted from MP. Cluster formula are cleaned up to match the notation used in literature and not reduced to indicate the actual number of atoms in the cluster. Names are extracted from literature. Asterisk refers to materials that were determined to be metastable. Melting point (mp.) and boiling point (bp.) are extracted from literature, as indicated by the source. The corresponding Wikipedia entry can be found using the name.

| Name | Formula | Cluster | MP-ID | $E_{\text{hull}}$ | $E_g$ | $\varepsilon_0$ | mp. | bp. | source |
|---|---|---|---|---|---|---|---|---|---|
| aluminum borohydride | Al(BH4)3 | Al(BH4)3 | mp-569787 | 0.0 | 6.24 | 3.40 | -64.5 | 44.5 | Wikipedia |
|  | Al2PdCl8 | Al2Cl8Pd | mp-27452 | 0.0 | 1.68 | 3.51 |  |  | [1] |
| aluminium iodide | AlI3 | Al2I6 | mp-30930 | 0.0 | 3.27 | 3.73 | 188.3 | 382.0 | Wikipedia |
|  | AlSCl3O2 | [Cl2Al(mu-O2SCl)]2 | mp-556418 | 0.0 | 3.98 | 3.51 | 0.0 |  | [2] |
| arsenic trioxide (cubic) | As2O3 | As4O6 | mp-2184 | 0.009 | 4.03 | 3.83 | 312.2 | 465.0 | Wikipedia |
| uzonite | As4S5 | As4S5 | mp-502 | 0.003 | 2.01 | 5.45 |  |  | [3] |
| arsenic trichloride | AsCl3 | AsCl3 | mp-23280 | 0.0 | 4.10 | 3.33 | -16.2 | 130.2 | Wikipedia |
|  | AsCl3O | (AsOCl3)2 | mp-29863 | 0.0 | 2.54 | 3.13 | 0.0 |  | [4] |
| arsenic pentachloride | AsCl5 | AsCl5 | mp-30106 | 0.0 | 1.46 | 3.07 | -50.0 |  | Wikipedia |
| arsenic trifluoride | AsF3 | AsF3 | mp-28027 | 0.0 | 5.40 | 5.24 | -8.5 | 60.4 | Wikipedia |
| arsenic pentafluoride | AsF5 | AsF5 | mp-8723 | 0.0 | 4.63 | 2.35 | -79.8 | -52.8 | Wikipedia |
| realgar | AsS | As4S4 | mp-542846 | 0.004 | 2.05 | 4.85 | 320.0 |  | Wikipedia |
| gold chloride | AuCl3 | Au2Cl6 | mp-27647 | 0.0 | 1.39 | 3.93 | 254.0 |  | Wikipedia |
| boric acid | B(HO)3 | B(OH)3 | mp-759069 | 0.003 | 6.09 | 2.81 | 170.9 | 300.0 | Wikipedia |
| * | B3H5 | B6H10 | mp-29721 | 0.038 | 3.88 | 2.85 | -62.3 | 108.0 | [5] |
|  | BBr | B9Br9 | mp-685043 | 0.009 | 1.43 | 3.71 |  |  | [6] |
| boron tribromide | BBr3 | BBr3 | mp-23225 | 0.0 | 3.87 | 2.37 | -46.3 | 91.3 | Wikipedia |
| boron trichloride | BCl3 | BCl3 | mp-23184 | 0.0 | 4.75 | 2.20 | -107.3 | 12.6 | Wikipedia |
| boron trifluoride | BF3 | BF3 | mp-558149 | 0.0 | 8.16 | 2.19 | -126.8 | -100.3 | Wikipedia |
| * H2O-BF3 adduct | BH2OF3 | H2OBF3 | mp-707009 | 0.0 | 7.57 | 3.54 |  |  | Wikipedia |
| ammonia borane | BH6N | BNH6 | mp-35082 | 0.001 | 6.07 | 5.19 | 104.0 |  | Wikipedia |
| * | BH6N | BNH6 | mp-675418 | 0.012 | 5.89 | 9.14 |  |  | - |
| boron triiodide | BI3 | BI3 | mp-23189 | 0.0 | 0.00 | 2.94 | 49.9 | 210.0 | Wikipedia |
| dibromine trioxide | Br2O3 | Br2O3 | mp-28933 | 0.0 | 1.79 | 7.16 | -40.0 |  | Wikipedia |
| bromine pentafluoride | BrF5 | BrF5 | mp-27987 | 0.0 | 3.46 | 3.16 | -61.3 | 40.2 | Wikipedia |
| bromyl fluoride | BrO2F | BrFO2 | mp-36262 | 0.0 | 2.90 | 3.68 | -9.0 | 55.0 | [7] |
| dichlorodifluoromethane | C(ClF)2 | CCl2F2 | mp-22966 | 0.015 | 5.74 | 2.43 | -157.7 | -29.8 | Wikipedia |
| trichlorofluoromethane | CCl3F | CCl3F | mp-23071 | 0.012 | 4.90 | 3.15 | -110.5 | 23.8 | Wikipedia |
| chlorotrifluoromethane | CClF3 | CClF3 | mp-28473 | 0.012 | 6.72 | 31.05 | -181.0 | -81.5 | Wikipedia |
| tetrafluoromethane | CF4 | CF4 | mp-1167 | 0.0 | 9.07 | 1.93 | -183.6 | -127.8 | Wikipedia |
| cyanuric chloride | CNCl | (NCCl)3 | mp-571324 | 0.0 | 3.79 | 2.78 | 144.0 | 192.0 | Wikipedia |
| dichlorine heptoxide | Cl2O7 | Cl2O7 | mp-31050 | 0.0 | 3.68 | 3.14 | -91.6 | 82.0 | Wikipedia |
| chlorine trifluoride | ClF3 | ClF3 | mp-556767 | 0.005 | 2.63 | 3.52 | -76.3 | 11.8 | Wikipedia |
| CoBr2.4H2O | CoH8(BrO2)2 | CoBr2(H2O)4 | mp-23965 | 0.053 | 2.20 | 5.03 |  |  | [8, 9] |
| * chromium hexafluoride | CrF6 | CrF6 | mp-1539213 | 0.0 | 2.09 | 3.21 |  |  | Wikipedia |
| FeBr2.4H2O | FeH8(BrO2)2 | FeBr2(H2O)4 | mp-24537 | 0.026 | 3.72 | 4.96 |  |  | [9] |
| gallium bromide | GaBr3 | Ga2Br6 | mp-30953 | 0.0 | 3.36 | 3.37 | 121.5 | 278.8 | Wikipedia |
| gallium trichloride | GaCl3 | Ga2Cl6 | mp-30952 | 0.0 | 4.28 | 3.36 | 77.9 | 201.0 | Wikipedia |
| Hg2(GaCl4)2 | GaHgCl4 | Hg2(GaCl4)2 | mp-1103091 | 0.0 | 2.86 | 7.45 |  |  | [10] |
| germanium tetrachloride | GeCl4 | GeCl4 | mp-30086 | 0.0 | 4.27 | 2.56 | -49.5 | 86.5 | Wikipedia |
| germanium tetrafluoride | GeF4 | GeF4 | mp-9816 | 0.0 | 5.68 | 2.69 | -15.0 | -36.5 | Wikipedia |
| germyl chloride | GeH3Cl | GeH3Cl | mp-28369 | 0.0 | 4.76 | 3.78 | 0.0 |  | [11] |

| Name | Formula | Formula2 | MP ID | | | | | | Ref |
|---|---|---|---|---|---|---|---|---|---|
| hydrogen peroxide | H2O2 | H2O2 | mp-28015 | 0.082 | 4.51 | 3.95 | -0.4 | 150.2 | Wikipedia |
| sulfuric acid | H2SO4 | H2SO4 | mp-625475 | 0.0 | 6.09 | 3.77 | 10.3 | 337.0 | Wikipedia |
| * | H2SO4 | H2SO4 | mp-690733 | 0.000 | 6.11 | 3.59 | | | - |
| * | H2SO4 | H2SO4 | mp-24172 | 0.004 | 6.05 | 3.86 | | | - |
| selenous acid | H2SeO3 | H2SeO3 | mp-27996 | 0.003 | 4.50 | 5.83 | 70.0 | | Wikipedia |
| selenic acid | H2SeO4 | H2SeO4 | mp-23866 | 0.0 | 3.42 | 4.81 | 58.0 | 260.0 | Wikipedia |
| ammonia | H3N | H3N | mp-29145 | 0.0 | 4.34 | 2.99 | -77.7 | -33.3 | Wikipedia |
| * | H3N | H3N | mp-643432 | 0.017 | 4.46 | 3.52 | | | - |
| * | H3N | H3N | mp-779689 | 0.022 | 3.75 | 3.26 | | | - |
| methane | H4C | CH4 | mp-1021328 | 0.0 | 7.64 | 1.66 | -182.5 | -161.5 | Wikipedia |
| urea | H4CN2O | CH4N2O | mp-23778 | 0.0 | 5.08 | 3.80 | 133.0 | | Wikipedia |
| * | H4CN2O | CH4N2O | mp-976707 | 0.023 | 5.05 | 4.65 | | | - |
| thiourea | H4CSN2 | CH4N2S | mp-721896 | 0.006 | 3.33 | 7.41 | 182.0 | | Wikipedia |
| * | H4CSN2 | CH4N2S | mp-634059 | 0.054 | 2.84 | 5.54 | | | - |
| platinic acid | H8PtO6 | H2Pt(OH)6 | mp-625112 | 0.069 | 1.59 | 40.58 | | | [12] |
| * | H8PtO6 | H2Pt(OH)6 | mp-625111 | 0.070 | 1.60 | 11.82 | | | - |
| * | H8PtO6 | H2Pt(OH)6 | mp-625113 | 0.084 | 1.52 | 9.37 | | | - |
| iodic acid | HIO3 | HIO3 | mp-556216 | 0.018 | 3.07 | 8.64 | 110.0 | | Wikipedia |
| | HS8N | HNS8 | mp-29491 | 0.014 | 2.61 | 3.59 | 85.0 | | [13] |
| mercury bromide | HgBr | Hg2Br2 | mp-23177 | 0.0 | 2.46 | 25.24 | 405.0 | 390.0 | Wikipedia |
| mercury chloride | HgCl | Hg2Cl2 | mp-22897 | 0.0 | 2.82 | 17.40 | 276.0 | 304.0 | Wikipedia |
| iodine trichloride | ICl3 | I2Cl6 | mp-27729 | 0.0 | 1.81 | 4.71 | 63.0 | | Wikipedia |
| iodine heptafluoride | IF7 | IF7 | mp-27988 | 0.0 | 1.55 | 2.50 | 4.5 | 4.8 | Wikipedia |
| | LiS4 | Li2S8 | mp-995393 | 0.0 | 2.16 | 1.27 | | | No expt data found |
| magnesium perchlorate | MgH8(ClO6)2 | Mg(ClO4)2(H2O)4 | mp-865188 | 0.0 | 5.33 | 4.79 | 251.0 | | Wikipedia |
| * | MgH8(ClO6)2 | Mg(ClO4)2(H2O)4 | mp-989229 | 0.009 | 5.22 | 4.98 | | | - |
| molybdenum hexafluoride | MoF6 | MoF6 | mp-558836 | 0.005 | 4.25 | 2.55 | 17.5 | 34.0 | Wikipedia |
| chlorine nitrate | NClO3 | ClNO3 | mp-754712 | 0.0 | 2.94 | 3.10 | -107.0 | 18.0 | [14] |
| nitric oxide | NO2 | N2O4 | mp-557865 | 0.020 | 2.87 | 2.29 | -151.8 | -151.8 | [15] |
| | NbAlCl8 | AlCl8Nb | mp-28358 | 0.0 | 2.02 | 3.69 | 139.7 | | [16] |
| niobium iodide | NbI5 | Nb2I10 | mp-569578 | 0.0 | 0.99 | 8.83 | 543.0 | | Wikipedia |
| osmium tetroxide | OsO4 | OsO4 | mp-540783 | 0.0 | 3.30 | 2.57 | 40.2 | 129.7 | Wikipedia |
| * | OsO4 | OsO4 | mp-551905 | 0.001 | 3.29 | 2.65 | | | - |
| osmium oxide pentafluoride | OsOF5 | F5OOs | mp-555514 | 0.000 | 0.81 | 3.42 | 32.5 | | [17] |
| phosphorus pentoxide | P2O5 | P4O10 | mp-562613 | 0.010 | 5.55 | 2.88 | 340.0 | 360.0 | Wikipedia |
| phosphorus sulfoxide | P2S2O3 | P4S4O6 | mp-3667 | 0.0 | 4.01 | 3.13 | 102.0 | 295.0 | [18] |
| Pt(PF3)4 | P4PtF12 | Pt(PF3)4 | mp-555863 | 0.0 | 4.99 | 2.91 | -15.0 | 86.0 | [19] |
| phosphorus sulfide | P4S5 | P4S5 | mp-690 | 0.004 | 2.49 | 4.32 | 288.0 | 514.0 | Wikipedia |
| phosphorus selenide | P4Se5 | P4Se5 | mp-2447 | 0.018 | 1.83 | 6.80 | | | Wikipedia |
| phosphorus tribromide | PBr3 | Br3P | mp-27257 | 0.0 | 3.35 | 3.03 | -41.5 | 173.2 | Wikipedia |
| phosphorus oxybromide | PBr3O | Br3OP | mp-558645 | 0.0 | 3.33 | 4.04 | 56.0 | 189.5 | [20] |
| | PBrNF | P3N3Br3F3 | mp-559366 | 0.014 | 4.25 | 3.08 | | | [21] |
| phosphorus trichloride | PCl3 | PCl3 | mp-23230 | 0.0 | 4.08 | 2.77 | -93.6 | 76.1 | Wikipedia |
| phosphoryl chloride | PCl3O | POCl3 | mp-27277 | 0.0 | 4.68 | 3.29 | 1.2 | 105.8 | Wikipedia |
| * | PCl3O | POCl3 | mp-753611 | 0.006 | 4.56 | 3.52 | | | - |
| phosphoryl chloride difluoride | PClOF2 | ClF2OP | mp-558681 | 0.0 | 5.73 | 2.88 | -96.4 | 3.1 | Wikipedia |
| phosphorus pentafluoride | PF5 | PF5 | mp-8511 | 0.0 | 7.11 | 2.15 | -93.8 | -84.6 | Wikipedia |
| fluorophosphorane | PH2F3 | PH2F3 | mp-29515 | 0.0 | 6.47 | 12.44 | -47.0 | 0.9 | [22] |
| | PH3CS3 | (CH3S)2(P2S2)S2 | mp-559616 | 0.021 | 2.18 | 4.21 | | | [23] |
| NH3-PF5 adduct | PH3NF5 | NH3PF5 | mp-722832 | 0.0 | 7.09 | 6.27 | | | [24] |
| phosporic acid | PH3O4 | H3PO4 | mp-23902 | 0.0 | 6.23 | 4.76 | 40.0 | 212.0 | Wikipedia |
| * | PH3O4 | H3PO4 | mp-626449 | 0.011 | 5.81 | 5.91 | | | - |
| * | PH3O4 | H3PO4 | mp-626464 | 0.023 | 5.65 | 5.53 | | | - |
| | PHF4 | PHF4 | mp-29514 | 0.0 | 7.55 | 2.59 | -80.0 | -39.0 | [25] |
| hydrophosphoryl difluoride | PHOF2 | OPF2H | mp-698060 | 0.011 | 7.00 | 3.22 | 55.0 | 55.0 | [26] |
| hydrothiophosphoryl difluoride | PHSF2 | SPF2H | mp-642795 | 0.0 | 4.57 | 3.14 | 0.2 | 0.2 | [26] |
| phosphorus iodide | PI2 | P2I4 | mp-29443 | 0.0 | 2.06 | 4.25 | 125.5 | | Wikipedia |
| phosphorus iodide | PI3 | PI3 | mp-27529 | 0.0 | 2.36 | 4.11 | 61.2 | 200.0 | Wikipedia |
| hexachlorophosphazene | PNCl2 | (NPCl2)3 | mp-571213 | 0.000 | 3.85 | 3.24 | 112.0 | | Wikipedia |
| | PNClF | P3N3Cl3F3 | mp-554472 | 0.008 | 5.08 | 2.89 | | | [21] |
| (PNF2)4 | PNF2 | F8N4P4 | mp-555292 | 0.004 | 5.17 | 3.28 | 28.0 | 89.7 | [27] |
| | Pd(SCl3)2 | PdS2Cl6 | mp-28174 | 0.0 | 1.62 | 5.32 | | | [28] |
| | Pd(SeCl3)2 | PdSe2Cl6 | mp-28175 | 0.0 | 1.72 | 4.67 | | | [29] |
| palladium chloride | PdCl2 | Pd6Cl12 | mp-29487 | 0.0 | 1.43 | 3.96 | 679.0 | | Wikipedia |

| Name | Formula | Cluster | MP-ID | $E_{\text{hull}}$ | $E_g$ | $\varepsilon_0$ | mp. | bp. | source |
|---|---|---|---|---|---|---|---|---|---|
| platinum chloride | PtCl2 | Pt6Cl12 | mp-23290 | 0.0 | 1.83 | 3.29 | 581.0 | | Wikipedia |
| perrhenic acid | Re2H4O9 | Re2O7(OH2)2 | mp-625238 | 0.026 | 2.64 | 11.38 | 115.0 | | [30] |
| * | Re2O7 | Re2O7 | mvc-7040 | 0.039 | 3.47 | 3.19 | | | - |
| ruthenium tetroxide | RuO4 | RuO4 | mp-554791 | 0.0 | 2.43 | 2.77 | 25.4 | 40.0 | Wikipedia |
| | RuS3Cl8 | Ru2S6Cl16 | mp-29568 | 0.0 | 1.90 | 5.24 | 122.0 | | [31] |
| sulfur (alpha) | S | S8 | mp-77 | 0.001 | 2.71 | 2.95 | 115.2 | 444.6 | Wikipedia |
| * | S | S12 | mp-558014 | 0.014 | 2.48 | 3.09 | | | - |
| * | S | S14 | mp-561513 | 0.019 | 2.43 | 3.44 | | | - |
| sulfuryl chloride | S(ClO)2 | SO2Cl2 | mp-28405 | 0.0 | 3.79 | 3.46 | -54.1 | 69.4 | Wikipedia |
| sulfuryl trifluoride | S(OF)2 | SO2F2 | mp-8537 | 0.0 | 5.90 | 2.33 | -124.7 | -55.4 | Wikipedia |
| disulfuryl difluoride | S2O5F2 | S2O5F2 | mp-28676 | 0.0 | 5.61 | 2.75 | -48.1 | 50.9 | [32] |
| disulfur dibromide | SBr | Br2S2 | mp-28099 | 0.0 | 2.31 | 5.41 | 46.0 | 46.0 | Wikipedia |
| thionyl bromide | SBr2O | Br2OS | mp-28407 | 0.0 | 2.73 | 3.57 | 0.0 | | [33] |
| disulfur dichloride | SCl | S2Cl2 | mp-28096 | 0.0 | 2.89 | 3.22 | -80.0 | 137.1 | Wikipedia |
| thionyl chloride | SCl2O | SOCl2 | mp-28406 | 0.0 | 3.64 | 3.05 | -104.5 | 74.6 | Wikipedia |
| sulfuryl chloride fluoride | SClO2F | SO2ClF | mp-554012 | 0.0 | 4.68 | 2.79 | -124.7 | 7.1 | Wikipedia |
| sulfur hexafluoride | SF6 | SF6 | mp-8560 | 0.0 | 5.91 | 2.15 | -64.0 | -50.8 | Wikipedia |
| * | SF6 | SF6 | mp-975 | 0.001 | 5.92 | 2.38 | | | - |
| sulfur trioxide | SO3 | S3O9 | mp-2414 | 0.0 | 5.14 | 2.78 | 16.9 | 45.0 | Wikipedia |
| antimony trioxide | Sb2O3 | Sb4O6 | mp-1999 | 0.008 | 3.41 | 8.42 | 656.0 | 1425.0 | Wikipedia |
| antimony tribromide | SbBr3 | Br3Sb | mp-27399 | 0.0 | 3.45 | 4.00 | 96.6 | 288.0 | Wikipedia |
| antimony trichloride | SbCl3 | SbCl3 | mp-22872 | 0.0 | 3.75 | 7.67 | 73.4 | 223.5 | Wikipedia |
| | SbCl3F2 | (SbCl3F2)4 | mp-560748 | 0.004 | 1.71 | 4.47 | 62.0 | | [34] |
| selenium (beta) | Se | Se8 | mp-542605 | 0.006 | 1.40 | 5.33 | 221.0 | 685.0 | Wikipedia |
| diselendibromid | SeBr | Br2Se2 | mp-570589 | 0.006 | 1.55 | 6.31 | -49.0 | 225.0 | [35] |
| selenium tetrafluoride | SeF4 | SeF4 | mp-29172 | 0.0 | 4.14 | 6.68 | -13.2 | 101.0 | Wikipedia |
| seleninyl difluoride | SeOF2 | F2OSe | mp-27367 | 0.0 | 4.10 | 4.18 | 15.0 | | [36] |
| | Si2Cl2O3 | Cl8Si8O12 | mp-28959 | 0.026 | 5.68 | 2.94 | | | [37] |
| octahydridosilsesquioxane | Si2H2O3 | Si8H8O12 | mp-24431 | 0.023 | 6.37 | 3.23 | | | [38] |
| | Si2H2S3 | Si4H4S6 | mp-28090 | 0.0 | 3.37 | 3.62 | | | [39] |
| disiloxane | Si2H6O | Si2H6O | mp-27949 | 0.050 | 5.26 | 2.86 | -15.2 | -15.2 | Wikipedia |
| disilyl selenide | Si2H6Se | Si2H6Se | mp-29310 | 0.000 | 3.75 | 3.62 | -68.0 | 85.0 | [40] |
| silicon tetrabromide | SiBr4 | SiBr4 | mp-570285 | 0.0 | 4.11 | 2.61 | 5.0 | 153.0 | Wikipedia |
| cyclic chlorosiloxane | SiCl2O | (SiCl2O)4 | mp-23079 | 0.015 | 5.57 | 2.81 | 77.0 | | [41] |
| cyclic chlorosiloxane | SiCl2O | (SiCl2O)3 | mp-23563 | 0.027 | 5.57 | 2.64 | 43.5 | | [41] |
| silicon tetrachloride | SiCl4 | SiCl4 | mp-28391 | 0.0 | 5.79 | 2.40 | -68.7 | 57.6 | Wikipedia |
| silicon tetrafluoride | SiF4 | SiF4 | mp-1818 | 0.0 | 7.94 | 2.16 | -95.0 | -90.3 | Wikipedia |
| silyl fluoride | SiH3F | SiH3F | mp-28289 | 0.024 | 6.01 | 2.62 | -98.6 | -98.6 | [42] |
| iodosilane | SiH3I | SiH3I | mp-28538 | 0.0 | 4.21 | 3.70 | -57.0 | 45.4 | [43] |
| silane | SiH4 | SiH4 | mp-23739 | 0.005 | 6.59 | 2.47 | -185.0 | -111.9 | Wikipedia |
| SiF4-2NH3 adduct | SiH6(NF2)2 | SiF4(NH3)2 | mp-643277 | 0.0 | 6.70 | 3.21 | 166.0 | | [44] |
| tin iodide | SnI4 | SnI4 | mp-571436 | 0.019 | 1.78 | 3.40 | 143.0 | 348.5 | Wikipedia |
| | Ta4S9Br8 | Ta4S9Br8 | mp-559724 | 0.0 | 1.30 | 4.38 | | | [45] |
| | TaCl4F | (TaCl4F)4 | mp-27854 | 0.0 | 3.10 | 4.44 | | | [46] |
| telluric acid | Te(HO)6 | Te(OH)6 | mp-626012 | 0.050 | 2.23 | 5.28 | 136.0 | | Wikipedia |
| * | Te(HO)6 | Te(OH)6 | mp-625993 | 0.060 | 2.00 | 5.40 | | | - |
| trioxotetrafluoroditelluric acid | Te2H2O3F4 | H2Te2O3F4 | mp-1191341 | 0.0 | 4.62 | 7.89 | | | [47] |
| tellurium hexafluoride | TeF6 | TeF6 | mp-1875 | 0.0 | 4.40 | 2.47 | -38.9 | -37.6 | Wikipedia |
| * | TeF6 | TeF6 | mp-1542727 | 0.001 | 4.44 | 2.44 | | | - |
| | TeS7Br2 | Br2S7Te | mp-683875 | 0.003 | 2.23 | 4.12 | | | [48] |
| | TeS7Cl2 | Cl2S7Te | mp-672186 | 0.0 | 2.26 | 4.43 | | | [48] |
| titanium tetrabromide | TiBr4 | TiBr4 | mp-569814 | 0.0 | 2.84 | 2.78 | 39.0 | 230.0 | Wikipedia |
| titanium tetrachloride | TiCl4 | TiCl4 | mp-30092 | 0.0 | 3.47 | 2.52 | -24.1 | 136.4 | Wikipedia |
| tungsten hexachloride | WCl6 | WCl6 | mp-571518 | 0.0 | 1.92 | 4.94 | 275.0 | 346.7 | Wikipedia |
| * | WCl6 | WCl6 | mp-23178 | 0.000 | 1.75 | 4.45 | | | - |
| xenon tetrafluoride | XeF4 | XeF4 | mp-23185 | 0.0 | 2.86 | 3.48 | 117.0 | | Wikipedia |

TABLE II: Additional materials found by graph-theoretical approach. All information is as in Table S1.

| Name | Formula | Cluster | MP-ID | $E_{\text{hull}}$ | $E_g$ | $\varepsilon_0$ | mp. | bp. | source |
|---|---|---|---|---|---|---|---|---|---|
| * bismuth oxide | Bi2O3 | Bi4O6 | mp-1188267 | 0.072 | 2.60 | | | | Wikipedia |
| * nitrogen | N2 | N10 | mp-1080711 | 0.0 | 0.00 | | | | [49] |

| | | | | | | | | |
|---|---|---|---|---|---|---|---|---|
| phosphorus trioxide | P2O3 | P4O6 | mp-368 | 0.088 | 3.85 | 23.8 | 173.1 | Wikipedia |
| P4S5 (beta) | P4S5 | P4S5 | mp-7260 | 0.017 | 2.61 | 103.0 | | [50] |
| arsenic selenide | AsSe | As4Se4 | mp-542570 | 0.005 | 1.54 | 265.0 | | [51] |
| P4S4 (alpha) | PS | P4S4 | mp-612 | 0.020 | 2.52 | 134.0 | | [50] |
| germanium thiobromide | Ge2S3Br2 | Ge4S6Br4 | mp-540792 | 0.0 | 2.46 | 305.0 | | [52] |
| germanium thioiodide | Ge2S3I2 | Ge4S6I4 | mp-27928 | 0.0 | 2.25 | 310.0 | | [52] |
| phosphorus pentasulfide | P2S5 | P4S10 | mp-541788 | 0.0 | 2.58 | 288.0 | 514.0 | Wikipedia |
| niobium bromide | NbBr5 | Nb2Br10 | mp-28601 | 0.0 | 1.56 | 254.0 | 364.0 | Wikipedia |
| * | NbBr5 | Nb2Br10 | mp-568245 | 0.004 | 1.60 | | | - |
| tantalum bromide | TaBr5 | Ta2Br10 | mp-1203015 | 0.0 | 2.03 | 265.0 | 349.0 | Wikipedia |
| * | TaBr5 | Ta2Br10 | mp-568846 | 0.000 | 2.03 | | | - |
| tungsten bromide | WBr5 | W2Br10 | mp-29554 | 0.012 | 0.00 | 286.0 | 333.0 | Wikipedia |
| * | WBr5 | W2Br10 | mp-1095321 | 0.012 | 0.00 | | | - |
| molybdenum chloride | MoCl5 | Mo2Cl10 | mp-569855 | 0.0 | 0.00 | 194.0 | 268.0 | Wikipedia |
| * | MoCl5 | Mo2Cl10 | mp-569436 | 0.003 | 0.00 | | | - |
| * | MoCl5 | Mo2Cl10 | mp-571256 | 0.003 | 0.00 | | | - |
| * | MoCl5 | Mo2Cl10 | mp-29147 | 0.019 | 0.00 | | | - |
| niobium chloride | NbCl5 | Nb2Cl10 | mp-23307 | 0.0 | 2.22 | 204.7 | 248.2 | Wikipedia |
| * | NbCl5 | Nb2Cl10 | mp-568483 | 0.001 | 2.30 | | | - |
| rhenium chloride | ReCl5 | Re2Cl10 | mp-27645 | 0.054 | 0.00 | 220.0 | | Wikipedia |
| antimony chloride | SbCl5 | Sb2Cl10 | mp-570796 | 0.003 | 1.85 | -54.1 | | [53] |
| vanadium chloride | VCl5 | V2Cl10 | mp-1101909 | 0.0 | 1.05 | -10.0 | | [54] |
| tungsten chloride | WCl5 | W2Cl10 | mp-27160 | 0.022 | 0.00 | 248.0 | 275.6 | Wikipedia |
| rhenium trichloride dioxide | ReCl3O2 | Re2O4Cl6 | mp-607436 | 0.0 | 1.87 | 35.0 | | [55] |
| tantalum iodide | TaI5 | Ta2I10 | mp-570679 | 0.0 | 1.17 | 382.0 | | Wikipedia |

TABLE III: Additional properties calculated using vdW-DF functional: Cluster formula, MP-ID, binding energy per cluster (eV), volume per cluster ($\text{Å}^3$), HOMO and LUMO levels of cluster aligned w.r.t. vacuum level (eV), and dipole ($e\text{Å}$).

| formula | MP-ID | binding energy | volume | HOMO | LUMO | dipole |
|---|---|---|---|---|---|---|
| Sb4O6 | mp-1999 | -4.52 | 170.94 | 2.92 | 4.09 | 0.01 |
| As4S5 | mp-502 | -3.31 | 222.33 | 1.42 | 2.64 | 0.13 |
| As4S4 | mp-542846 | -2.82 | 195.84 | 1.38 | 2.80 | 0.00 |
| As4O6 | mp-2184 | -2.63 | 162.33 | 4.04 | 4.49 | 0.00 |
| P4O10 | mp-562613 | -2.17 | 193.75 | 5.47 | 6.01 | 0.00 |
| Pt6Cl12 | mp-23290 | -1.69 | 427.50 | 1.48 | 2.01 | 0.00 |
| Pd6Cl12 | mp-29487 | -1.75 | 415.68 | 1.09 | 1.64 | 0.00 |
| Nb2I10 | mp-569578 | -2.30 | 454.88 | 0.61 | 1.08 | 0.00 |
| As4Se4 | mp-542570 | -6.78 | 208.38 | 0.85 | 2.17 | 0.00 |
| P4S10 | mp-541788 | -2.94 | 346.68 | 2.17 | 2.97 | 0.00 |
| Ge4S6Br4 | mp-540792 | -2.95 | 387.29 | 2.14 | 2.79 | 0.01 |
| Ge4S6I4 | mp-27928 | -3.15 | 455.05 | 2.06 | 2.59 | 0.02 |
| Nb2Br10 | mp-28601 | -6.58 | 368.70 | 1.37 | 1.70 | 0.00 |
| Ta2Br10 | mp-1203015 | -4.74 | 372.00 | 1.86 | 2.19 | 0.00 |
| Nb2Cl10 | mp-23307 | -3.87 | 315.41 | 2.07 | 2.33 | 0.00 |
| Ta2I10 | mp-570679 | -8.72 | 459.17 | 0.94 | 1.41 | 0.00 |

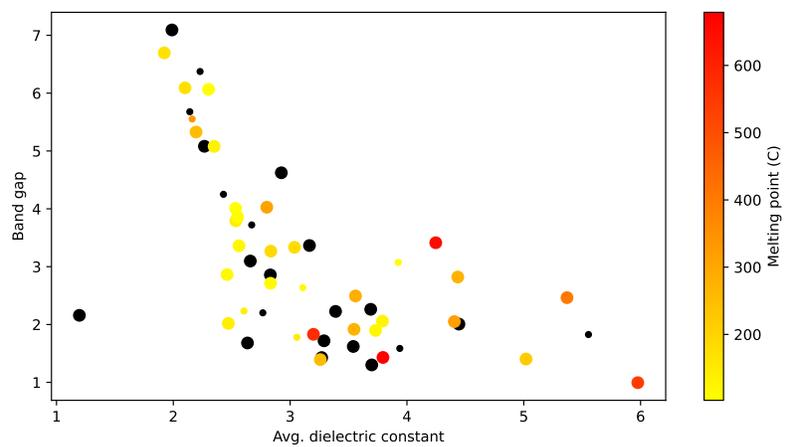

FIG. 1: Band gap vs. high-frequency dielectric constant $\varepsilon_\infty$. Otherwise the same as Figure 1(b) in the main paper.

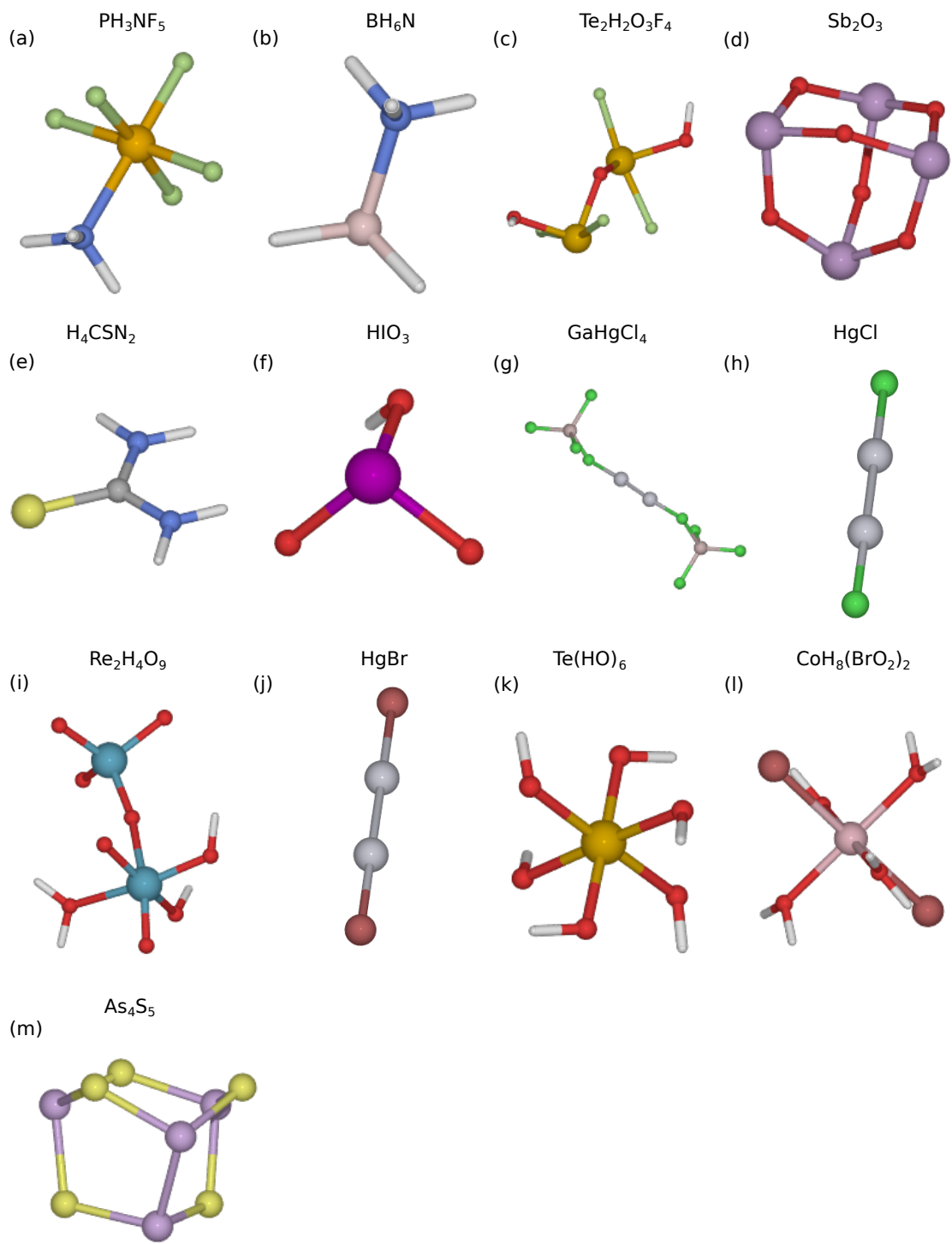

FIG. 2: Atomic structures for all materials included in Figure 1(c) of the main paper.

7

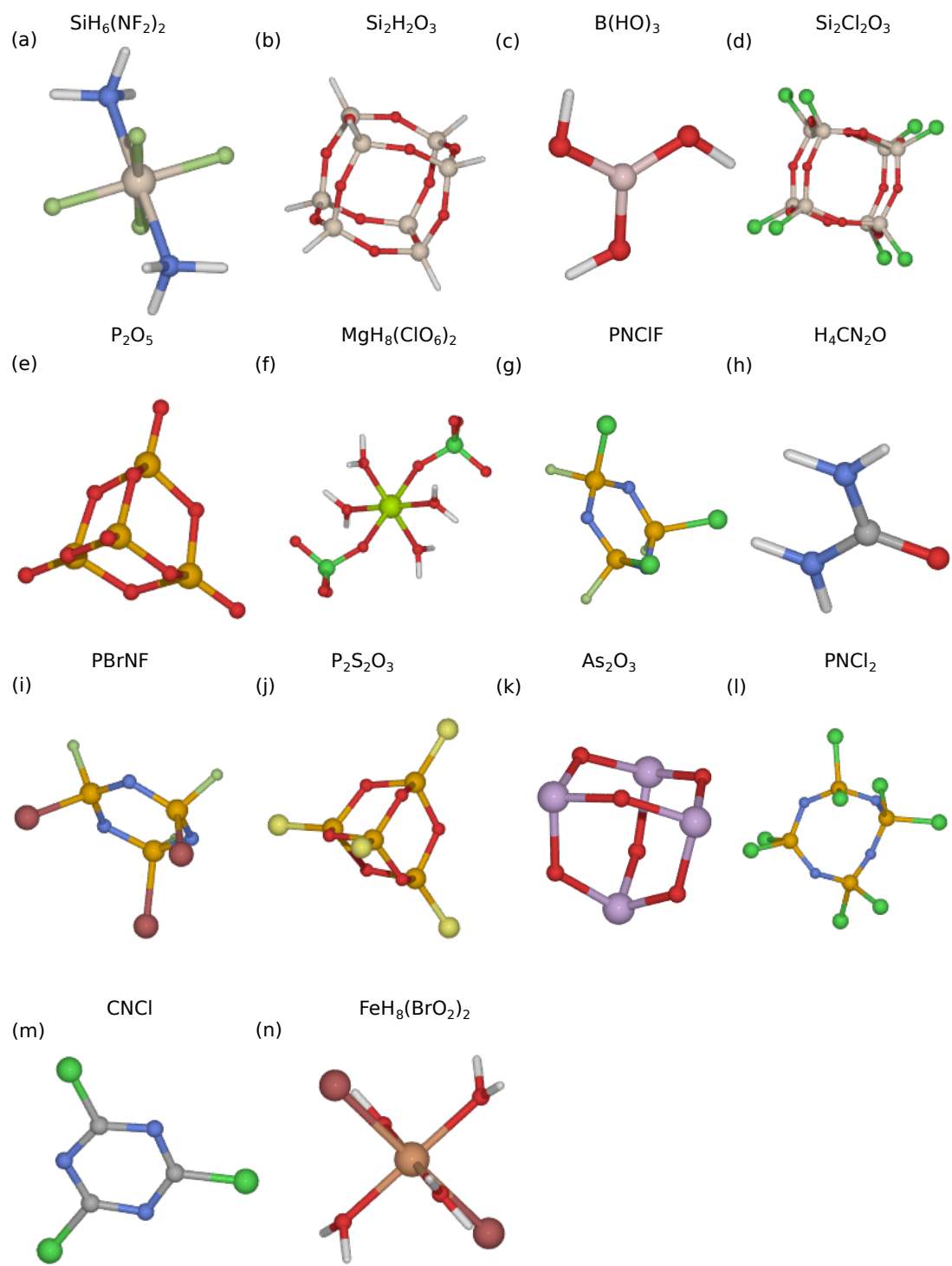

FIG. 3: Atomic structures for all materials included in Figure 1(d) of the main paper.

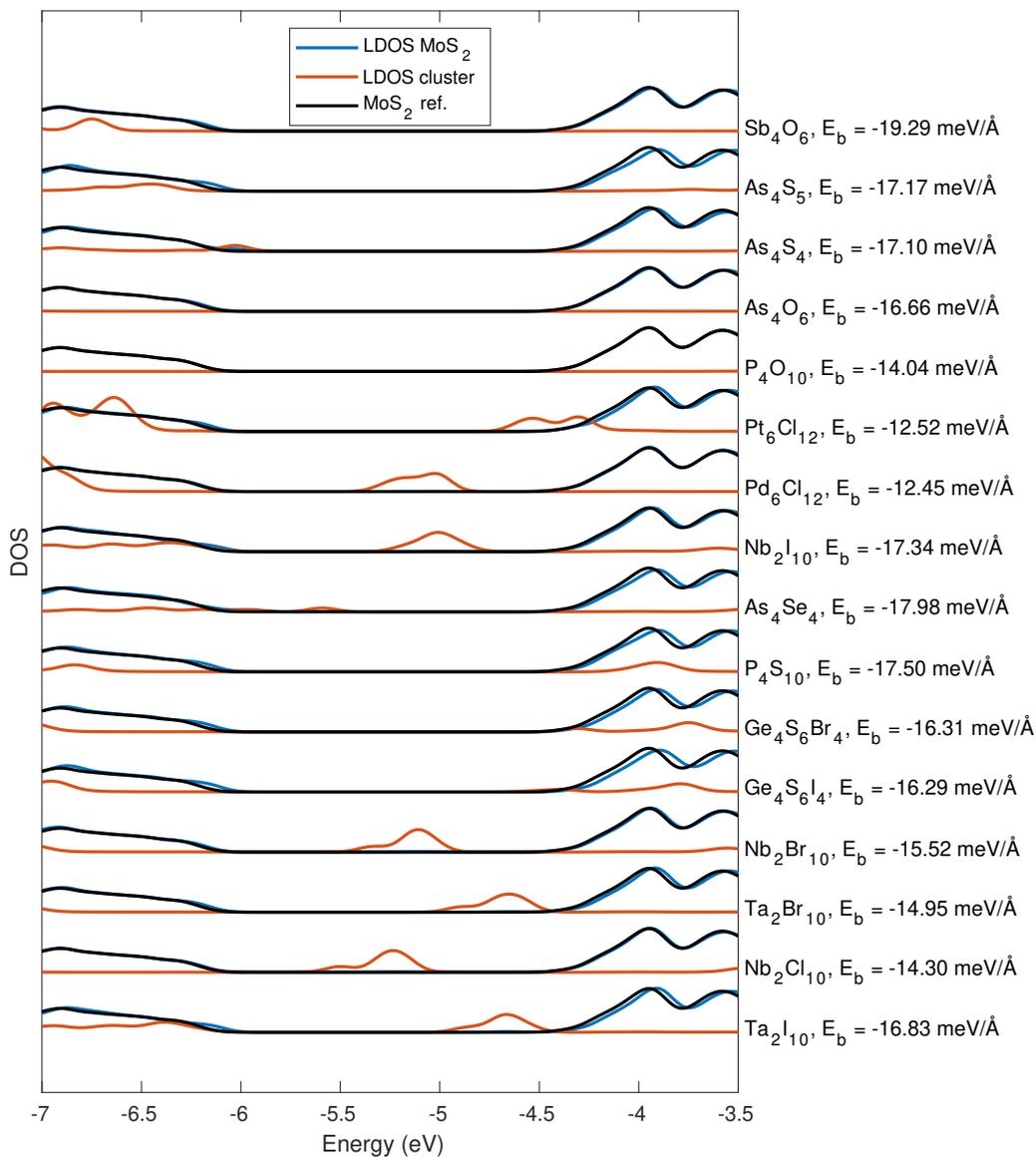

FIG. 4: Electronic structure of 0D clusters adsorbed on MoS$_2$. Local density of states of the cluster and of MoS$_2$ layer. Total DOS pristine MoS$_2$ is given as a reference. All DOS are aligned using the vacuum level.



\end{document}